\documentclass[prc,twocolumn]{revtex4}
\usepackage{amsmath,amssymb,epsfig,bm}
\newcommand{\be}{\begin{equation}}
\newcommand{\ee}{\end{equation}}
\newcommand{\bea}{\begin{eqnarray}}
\newcommand{\eea}{\end{eqnarray}}

\begin{document}
\title{Light nuclei in supernova envelopes: 
a quasiparticle gas model}

\author{Stefan Heckel, Philipp P. Schneider,  Armen~Sedrakian}
\address{Institute for Theoretical Physics, J. W. Goethe-University, 
D-60438 Frankfurt am Main, Germany}

\begin{abstract}
We present an equation of state and the composition of low-density supernova 
matter composed of light nuclei with mass number $A \le 13$. We work 
within the quasiparticle gas model, which accounts for bound states with decay 
time scales larger than the relevant time scale of supernova and protoneutron 
star evolution. The mean-field contribution is included in terms of Skyrme 
density functional. Deuterons, tritons, and $^3$H(e) nuclei appear in matter 
in concentrations that are substantially higher than those of heavier nuclei.
We calculate the critical temperature of deuteron condensation in such matter,
and demonstrate that the appearance of clusters substantially lowers the 
critical temperature.
\end{abstract}
\date{\today}

              
\pacs{97.60.Jd,26.60.+c,21.65.+f,13.15.+g}
\keywords{Low density nuclear matter, equation of state, clustering, 
          Bose-Einstein condensation}

\maketitle

\section{Introduction}

A key ingredient of studies of the formation of neutrino signal in 
supernova explosions and the supernova mechanism itself is the equation 
of state and composition of matter at the densities  $\rho \le 10^{12}$ 
g cm$^{-3}$ and within the temperature range $0\le T\le 10$ MeV~\cite{Bethe:1990mw}. 
The matter at densities below the nuclear saturation density is composed 
of a mixture of nuclei and free (unbound) nucleons with a charge neutralizing 
background of electrons.  The state of the art equations of state
that are routinely used in the current supernova simulations, 
include the nucleons, the $\alpha$ particles, and a heavy nucleus as the 
independent degrees of freedom~\cite{Lattimer:1991nc,Shen:1998gq}. 

Numerical simulations of core collapse supernovas demonstrate that 
about $10^{53}$ ergs of gravitational binding energy is liberated 
in neutrinos of all 
flavors~\cite{Burrows:2006ci,Janka:2006fh,Liebendorfer:2008}.
Over the time scales relevant for supernovas and protoneutron 
stars matter is opaque to neutrinos above 
densities $10^{11}$-$10^{12}$ g cm$^{-3}$. The last-scattering 
surface, known as the ``neutrinosphere,'' generates the neutrino 
spectrum of the supernova, which is potentially observable by the 
supernova neutrino detectors. The signal carries an imprint of 
physical conditions at the neutrinosphere and can provide information 
on properties of matter under supernova conditions and dense matter
in general~\cite{Lattimer:2006xb,WEBER_BOOK,Sedrakian:2006mq}. 
Neutrino interactions 
at the neutrinosphere are also of importance for setting the initial 
conditions for possible nucleosynthesis process in supernova winds.

In this work we focus our attention primarily on the composition 
and the equation of state of dilute isospin symmetric and asymmetric 
nuclear matter. Our goal is to introduce a simple setup for treating 
increasingly complex many-body problems related to light nuclei in 
supernova and protoneutron star matter. Our motivation lies 
in the Bose-Einstein condensation (BEC) of 
deuterons~\cite{Lombardo:2001ek,Sedrakian:2005db} 
and $\alpha$ particles~\cite{Sedrakian:2004fh,Funaki:2008rt}
in nuclear matter and  under supernova conditions~\cite{Sedrakian:2006xm}.
Furthermore, 
recent computations of two- and three-body binding energies in nuclear 
medium at nonzero temperature and density~\cite{Sedrakian:2005db}
allow us a fully quantum mechanical assessment of these effects beyond 
the occupied volume approximation. These goals are accomplished 
by the Quasiparticle Gas Model (QGM), described in the following 
section, which treats the nuclei as stable (infinite lifetime) entities 
over the time scales relevant for the supernova and protoneutron star
evolution. Our setup is sufficiently general to allow for resonant 
states (finite lifetime effects), the degeneracy of species, and hence,
the possibility of boson condensation and medium modifications of bound 
and scattering states in a clustered environment. The binding energies
of the nuclei $A\le 13$ can be computed from first principles in free 
space, e.g., within variational theory~\cite{Pieper:2001mp}. 
In this work we use the experimentally measured 
binding energies of light nuclei~\cite{Audi:2002rp}. 
Section~\ref{sec:conclusions} discusses the modifications of the 
binding energies of nuclei in the medium on the example of deuterons.
Our model includes the mandatory mean-field contribution to the energy 
density and the nucleon effective mass due to the momentum-dependent 
mean field (self-energy). 

In recent years a number of studies improved upon the equations of 
state and composition of Refs.~\cite{Lattimer:1991nc} and
\cite{Shen:1998gq}.
Instead of using one single heavy nucleus as a representative, 
an ensemble of nuclei with $A\le 1000$ was included in the
composition in Refs.~\cite{Souza:2008ud} and \cite{Botvina:2008su}, 
treating the nuclei 
as noninteracting Boltzmann gas. These statistical ensemble calculations
predict nuclei that are smaller than those obtained in a single 
(representative) nucleus approximation  and they find 
substantial amounts of light nuclei in the composition. The treatment 
of light nuclei has been 
improved by including the interactions among the nucleons and $\alpha$'s 
on the basis of phase shifts (virial expansion)~\cite{Horowitz:2005nd}. 
This approach has been extended further to include 
the three-body bound and scattering states~\cite{O'Connor:2007eb} 
and contributions of all nuclei up to $A\le 4$~\cite{Arcones:2008kv}. 
Neutrino interactions with light nuclei~\cite{Arcones:2008kv} and
the appearance of $A\le 4$ clusters in a dynamical 
simulation model~\cite{Sumiyoshi:2008qv}
have been explored. The changes in the binding energies of light nuclei 
within the Ritz variational 
theory are given in Ref.~\cite{Ropke:2008qk}. A different
view on clusterization in low-density nuclear matter emerges from the 
studies of the liquid-gas instability, which predicts clusterization 
of matter into fragments in the spinodal 
region~(Ref.~\cite{Ducoin:2008ic} and references therein).
The latter process is accessible in heavy-ion 
experiments~\cite{Shlomo:2009ny}.

The implications of the rich and complex composition of matter on 
the thermodynamics of matter and its effect on neutrino transport and 
other aspects of supernova physics are not yet fully understood. 
The purpose of this work is to advance the study of light 
clusters in supernovas in the following directions.
The composition of matter is extended to include all stable 
nuclei up to $A\le 13$, a number that is larger than that included in 
studies of light clusters to date.  In doing so the quantum statistics 
is fully included, i.e., the assumption of Boltzmann gas adopted, 
for example,
in nuclear statistical ensemble studies is relaxed. It follows 
then that any possible Bose-Einstein condensation (BEC) in clustered 
matter is automatically included in the theory. Indeed, we find that 
{\it there is a BEC of deuterons in the supernova environment}. The effect
of isospin asymmetry  on the composition of matter containing 
nuclei with mass number $A\le 4$ is studied. 
The formalism to address this issue is based on 
expressing the thermodynamical potential in terms of a sum over clusters, 
where each term is expressed through the {\it spectral function} 
of the corresponding cluster. This method, allows one 
to address a multitude of effects, such as finite decay width, 
short-lived states, Landau-Pomeranchuk suppression 
in radiation processes, etc. 

To summarize, the novelty of this work lies in the following:
first, the  formalism presented here has the advantage that it  
represents the contribution
of the clusters to the thermodynamic potential in terms of their 
{\it  spectral functions}. Second, we present a complete 
{\it quantum statistical} treatment of clusters up to 
$A \le 13$. Most previous 
work treats clusters as Boltzmann particles. Such an approach by default 
excludes any possible Bose-Einstein condensation. Furthermore, the 
majority of the previous works were restricted to clusters up to 
$A \le 4$, whereas we include clusters up to $A=13$. Third, we demonstrate
the {\it Bose-Einstein condensation of deuterons} in supernova matter.
Fourth, we demonstrate the dependence of the $A\le 4$ cluster abundances on 
{\it arbitrary isospin asymmetry}.

This article is organized as follows.  In Sec.~\ref{sec:QGM} we describe the 
quasiparticle model for a mixture of light nuclei in symmetric and 
asymmetric nuclear matter.  In Sec.~\ref{sec:results} we present the 
results for the composition, equation of state, and deuteron condensation
within the QGM. Section~\ref{sec:conclusions} studies the effects of
the in-medium modifications of the deuteron binding energies
and summarizes our results.

\section{Quasiparticle Gas Model}
\label{sec:QGM}

We consider matter composed of unbound nucleons and light nuclei with 
mass numbers $A\le 13$ in thermodynamical equilibrium at temperature 
$T$ and nucleon number density $n$. Each nucleus is characterized 
by its mass number $A$ and charge $Z$, which we collectively denote by 
$\alpha=(A,Z)$. We expand the thermodynamical potential of the system into 
a sum of contributions of clusters
\be\label{eq:thermopotential}
\Omega (\mu_n,\mu_p, T) =\sum_{\alpha}\Omega_{\alpha} (\mu_{\alpha}, T),
\ee
where $\mu_n$ and $\mu_p$ are the chemical potentials of neutrons and 
protons and $\mu_{\alpha}$ is the chemical potential of a nucleus, which 
is completely characterized by the variable $\alpha$. 
The chemical equilibrium among the species (baryon number and charge 
conservation) implies that
\be\label{eq:cheq}
\mu_{\alpha} = (A-Z)\mu_n + Z \mu_p.
\ee
At this stage one may either develop a direct perturbation theory 
for  the thermodynamic potential~\cite{Abrikosov:1963} or construct 
the Green's functions of the theory from appropriate Martin-Schwinger 
hierarchy~\cite{Sedrakian:2006mq} and express the thermodynamical 
potential in terms of densities. We follow the second path.
The Martin-Schwinger hierarchy is truncated with the help of self-energies 
such that the equation of motion for a nucleus $\alpha$ 
decouples from others. Then, the thermodynamic potential for each species 
is given by  
\be 
\Omega_{\alpha} (\mu, T)= - V\int_{-\infty}^{\mu_{\alpha}}
d\mu'_{\alpha}\,\, n_{\alpha}(\mu'_{\alpha}, T),
\ee
where $n_{\alpha}(\mu'_{\alpha}, T)$ is the number density 
of a nucleus  $\alpha =(A,Z)$. By introducing the Fourier 
transform of the finite temperature Green's function 
$iG^<_{\alpha}(x_1,x_2) = \langle \psi_{\alpha}(x_1)
\psi^{\dagger}_{\alpha}(x_2)\rangle$, where $\psi^{\dagger}_{\alpha}(x_1)$
and $\psi_{\alpha}(x_1)$ are the creation and annihilation operators of 
a nucleus $\alpha$ at the space-time point $x_1$, we write the densities 
of species as 
\bea\label{eq:density}
n_{\alpha} &=& ig_{\alpha}\int\frac{d\omega d^3p}{(2\pi)^4} 
G^<_{\alpha}(\omega,\vec p) \nonumber
\\ &=& g_{\alpha}\int\frac{d\omega d^3p}{(2\pi)^4}
S_{\alpha}(\omega,\vec p) f_{F/B}(\omega),
\eea
where $g_{\alpha}$ is the degeneracy factor for spin and isospin degrees
freedom and the Fermi/Bose distribution functions $f_{F/B}(\omega)$
account for statistical distribution of species with half-integer/integer 
spin,
\be 
f_{F/B}(\omega)=\left[1\pm {\rm exp}\left(\frac{\omega}{T}\right)
\right]^{-1}.
\ee
The spectral function is given by
\be 
S_{\alpha}(\omega,\vec p) = 
\frac{\Gamma_{\alpha}(\omega,\vec p)}
{[\omega-E_{\alpha}(\omega,\vec p)]^2
+\Gamma_{\alpha}^2(\omega,\vec p)/4},
\ee
where the quasiparticle energy is 
\be 
E_{\alpha}(\omega,\vec p) = \frac{p^2}{2Am}+B_{\alpha}+
{{\rm Re}\Sigma_{\alpha}(\omega,\vec p)}-\mu_{\alpha},
\ee
$B_{\alpha}$ is the binding energy of the nucleus 
$\alpha=(A,Z)$, $\Sigma_{\alpha}(\omega,\vec p)$ is 
its self-energy, and $\Gamma_{\alpha}(\omega,\vec p) 
= 2 {\rm Im}\Sigma_{\alpha}(\omega, \vec p)$ is the spectral 
width. We further assume that the nuclei under consideration 
are long-lived on the relevant time scales of supernova 
evolution, i.e., $\Gamma_{\alpha}(\omega,\vec p) = 0$. We also assume that the 
real parts of the self-energies are constants independent of momentum and 
frequency, in which case they can be absorbed in the chemical potential.
Finally, we neglect the effects of medium modification of binding 
energies; we return to this problem in the concluding section. Under 
these approximations, the spectral function is given by
\be 
S_{\alpha}(\omega,\vec p) = 2\pi\delta\left(\omega-\frac{p^2}{2Am}
-B_{\alpha}-{{\rm Re}\Sigma_{\alpha}}+\mu_{\alpha}
\right),
\ee
and the energy integral in Eq.~(\ref{eq:density}) is straightforward.
The defining feature of our model is now transparent -- the 
density is the sum of contributions from infinite lifetime 
quasiparticles (nuclei) characterized by the value $\alpha$. 
All relevant thermodynamic quantities can be computed from the 
thermodynamic potential Eq.~(\ref{eq:thermopotential}); 
the pressure and the entropy are given by
\begin{equation}
P= -\frac{\Omega}{V} , \quad \quad  
S = -\frac{\partial \Omega}{\partial T}.
\end{equation}
The pressure, entropy and other thermodynamical parameters 
of the electron gas are obtained from the thermodynamic potential
\bea\label{Omega_e}
\Omega_{e} = - g_eT\int\frac{d^3k}{(2\pi)^3}
{\rm ln}\left[f^{-1}\left(-E_{e}(k)+\mu_e\right)\right],
\eea
where electron degeneracy factor $g_e=2$, the electron energy is 
$E_{e} = \sqrt{k^2+m_e^2}$, where $m_e$ is the electron mass, 
and $\mu_e$ is the chemical potential. The electron density $n_e$ couples 
to the density of baryonic matter via the charge neutrality condition 
\be \label{charge_neutrality}
n_e - \sum_{\alpha} Zn_{\alpha} =0, 
\ee
where $n_e = \partial\Omega_e/\partial \mu_e$. The thermodynamical 
potential of positrons is obtained upon substituting $\mu_e\to -\mu_e$.
The thermodynamical potential of neutrinos of a given flavor has 
the same form as Eq.~(\ref{Omega_e}), where the neutrino mass 
and the chemical potential appear instead of  the electron ones and  
the neutrino degeneracy factor is $g_{\nu} =1$. The thermodynamical 
potential of antineutrinos is obtained in a similar fashion.

\section{Results}
\label{sec:results}

Under supernova conditions the electron fraction in matter 
is fixed and the evolution is nearly adiabatic (constant entropy). 
Here, to set the stage, we first explore the limit where the matter is 
isospin symmetric and isothermal. This discussion is followed by 
a study of a more general case of arbitrary isospin asymmetries.
Below, the isospin asymmetry is characterized either by the asymmetry
parameter $\chi = (n_n-n_p)/n$, where $n_n$, $n_p$ are the neutron and 
proton number densities and $n$ is the total number density or by the 
electron fraction $Y_e = n_e/n$ [see Eq.~(\ref{charge_neutrality})].
 Although large asymmetries are 
not realized in supernovas, a rapid neutronization process eventually 
equilibrates when the electron fraction reaches $Y_e \sim 0.05$ in 
protoneutron stars. 
The specific conditions prevailing in supernova matter, e.g.,
finite neutrino chemical potential, will be considered elsewhere.


\subsection{Density functional}
We start with a brief summary of the Skyrme density functional.
We assume that the nucleons interact via the Skyrme interaction, 
which is given by 
\begin{eqnarray}
 V(\vec r_1,\vec r_2) &=& t_0 \delta(\vec r_1 - \vec r_2) 
         + \frac{1}{2} t_1  \Bigl[ \delta(\vec r_1 - \vec r_2) 
         k^2  \nonumber \\ &+&
         k'^2 \delta(\vec r_1 - \vec r_2) \Bigr] 
        +t_2 \overleftarrow{k}' \delta(\vec r_1 - \vec r_2) 
\overrightarrow k \nonumber\\&+& \frac{1}{6} t_3 \rho 
\delta(\vec r_1 - \vec r_2), \label{e:skyrme_potential}
\end{eqnarray}
where  $\overrightarrow k = (\overrightarrow{\nabla}_1 
- \overrightarrow{\nabla}_2)/2i $ and $\overleftarrow{k}' 
= - (\overleftarrow{\nabla}_1 - \overleftarrow{\nabla}_2)/2i $, whereby
 $n$ is the density of nuclear matter. The parameters 
$t_0$, $t_1$, $t_2$ and $t_3$ are determined phenomenologically. 
We use the SkIII parametrization~\cite{Su:1987zz};
the parameter values are $t_0=-1128.75$ MeV fm$^3$, $t_1=395$ MeV fm$^5$, 
$t_2=-95$ MeV fm$^5$, and $t_3=1.4\times 10^{4}$ MeV fm$^6$. The 
(on-shell) quasiparticle spectrum for nucleons is given by 
$
\epsilon_i (p) = p^2/2m_i + {\rm Re}\Sigma (\epsilon_i (p), \vec p)-\mu_i  
$
which we take in the quasiparticle approximation, i.e., 
\be 
\epsilon_i (p) = \frac{p^2}{2m_i^*} - \mu'_i
\ee
where $i=n,p$ is the isospin index ($n$, neutrons; $p$, protons),  
$\mu_i' = \mu_i-{\rm Re}\Sigma (\epsilon_i (p_{F,i}), p_{F,i})$, where
$p_{F,i}$ is the Fermi momentum. The effective mass of a nucleon 
is computed from 
\begin{equation}
\frac{m_{n/p}^*}{m} =  \left[
1 + \frac{mn}{2}(t_1+t_2) +\frac{mn}{8}(t_2-t_1) (1\pm\chi) 
\right]^{-1}, \label{e:eff_mass}
\end{equation}
which we use as a correction for the masses of free {\it and} 
bound nucleons. The explicit form of the self-energy is immaterial, 
because its value at the Fermi surface can be absorbed in the 
chemical potential; we drop the prime on the chemical potentials 
hereafter.

\begin{figure}[t] 
\vskip 1.cm
\psfig{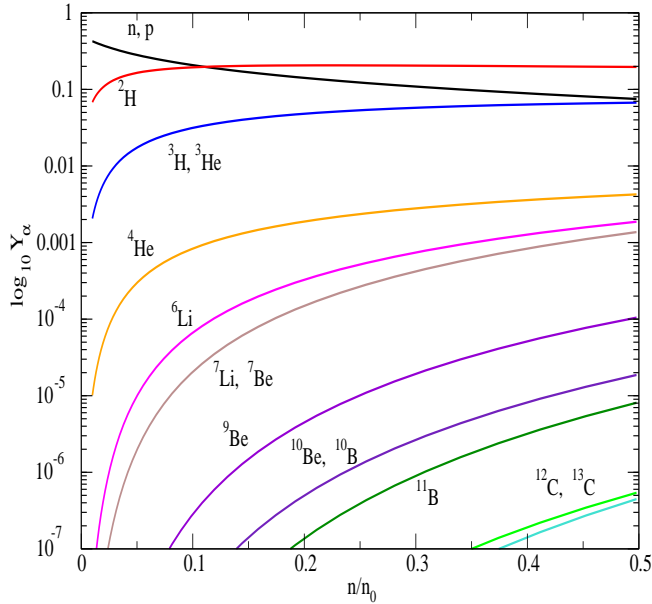}
\vskip 0.5cm
\caption{(Color online) Abundances of nuclei $Y_{\alpha} 
= n_{\alpha}/n$ in dilute isospin symmetrical matter composed of 
nuclei with mass numbers $A \le 13$ as a function of matter density, 
in units of nuclear saturation density  $n_0 = 0.16$ fm$^{-3}$, 
at $T=10$ MeV. The abundances of clusters decrease with increasing 
mass number.               
}\label{fig1}
\end{figure}
\begin{figure}[t]
\vskip 0.3cm 
\psfig{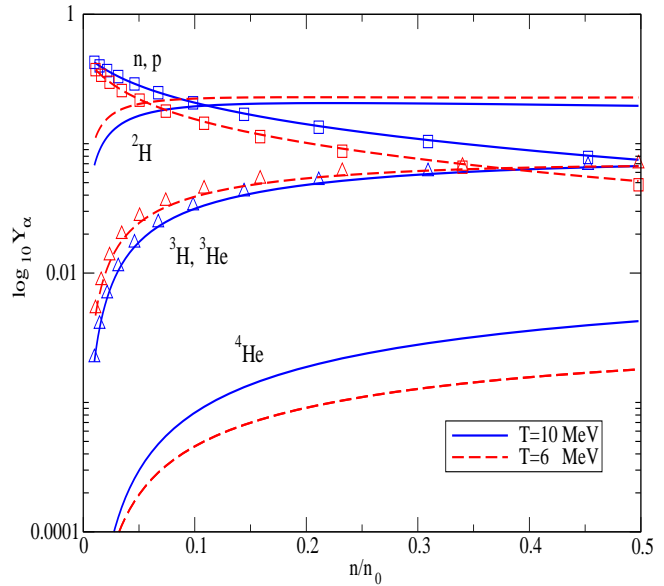}
\vskip 0.5cm
\caption{(Color online) Dependence of abundances of light nuclei $A\le 4$ on 
matter density, in units of $n_0$, for two temperatures $T=10$ MeV
(solid lines, blue online) and 6 MeV (dashed lines, red online).
Because of isospin symmetry, the abundances of protons and 
$^3$He nuclei, which are shown  by squares and triangles, respectively,
are nearly identical to those of neutrons and 
$^3$H nuclei.
}\label{fig2}
\end{figure}

\subsection{Isospin symmetric matter  ($\chi = 0$)}

Our numerical procedure uses the tabulated binding energies for 
nuclei with mass number $A\le 13$ with half-decay times that
are larger than the relevant dynamical time scales associated with 
supernova explosions~\cite{Audi:2002rp}. We first compute the effective 
masses of nucleons and the mean-field from the Skyrme density functional
with the SkIII parametrization. This is followed by a computation of 
the partial densities from Eq.~(\ref{eq:density}) with the normalization 
$n=\sum_{\alpha}n_{\alpha}$ and constraint (\ref{eq:cheq}), which 
provides us the chemical potentials of the species. Finally, 
we compute the thermodynamical potential (\ref{eq:thermopotential})
from which we obtain the pressure and the entropy. The effects of mean-field 
and mass renormalization are small at the relevant densities. Furthermore, 
the results shown below are insensitive to the choice of Skyrme 
parametrization.


Figure~\ref{fig1} displays the abundances of light nuclei, defined as
\be
Y_{\alpha} = \frac{n_{\alpha}}{n},
\ee
at constant
temperature $T=10$ MeV as a function of density (in units of nuclear 
saturation density $n_0 = 0.16$ fm$^{-3}$). At low densities 
the matter is dominated by nucleons with a small (about $10\%$)
admixture of deuterons. At intermediate densities the deuteron fraction 
becomes larger than that of the free nucleons; even though the 
population of  nuclei becomes more significant, those with $A\ge 4$ 
contribute less than $1\%$ to the total density. Next to deuterons
$^3$He and $^3$H nuclei are the dominant species in matter. 
The $\alpha$-particle abundance does not exceeds $0.5\%$ percent at any density. 
Medium modifications of binding energies 
of nuclei shift the balance between the abundances of nucleons 
and light nuclei in the high density part of Fig.~\ref{fig1}. Recall 
that as $n/n_0\to 1$ nuclei disappear asymptotically, leaving a continuum 
of nucleons. Note that numerically the abundances of neutrons and protons 
in mirror nuclei (obtained by an interchange of neutrons and protons) 
differ slightly because of the differences in their masses and binding 
energies; however, these differences are insignificant on the scales of 
the figure.

Figure~\ref{fig2} shows the abundances of dominant species for two 
different temperatures. Reducing the temperature 
from $T= 10 $ to $T=6$ MeV increases the 
abundances of light species, such as deuterons, $^3$H and $^3$He,
while  the abundance of $\alpha$ particles is suppressed
(note that here we assume that $B_{\alpha}=$ const., which 
is a valid assumption only in the low-density limit).

\begin{figure}[t] 
\vskip 1.cm
\psfig{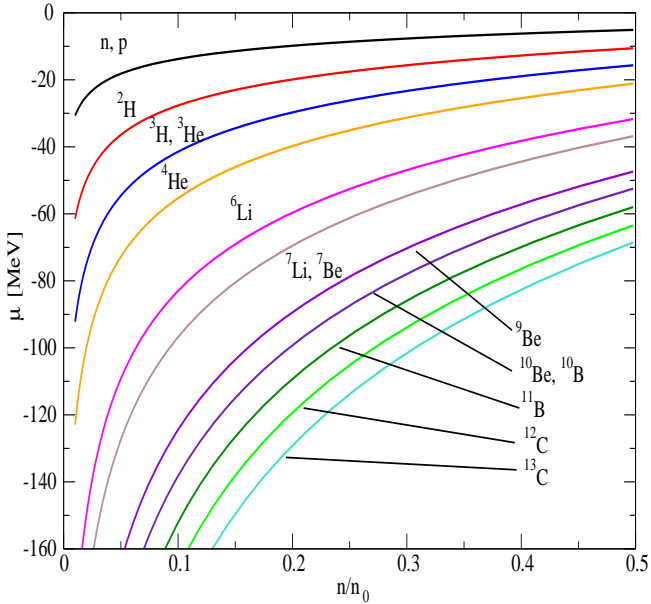}
\vskip 0.3cm
\caption{(Color online) Dependence of chemical potentials of 
$A \le 13$ mass number nuclei on matter density, 
in units of $n_0 = 0.16$ fm$^{-3}$, at $T=10$ MeV. The chemical 
potentials decrease with increasing mass number. 
}\label{fig3}
\end{figure}
\begin{figure}[t] 
\vskip 1.cm
\psfig{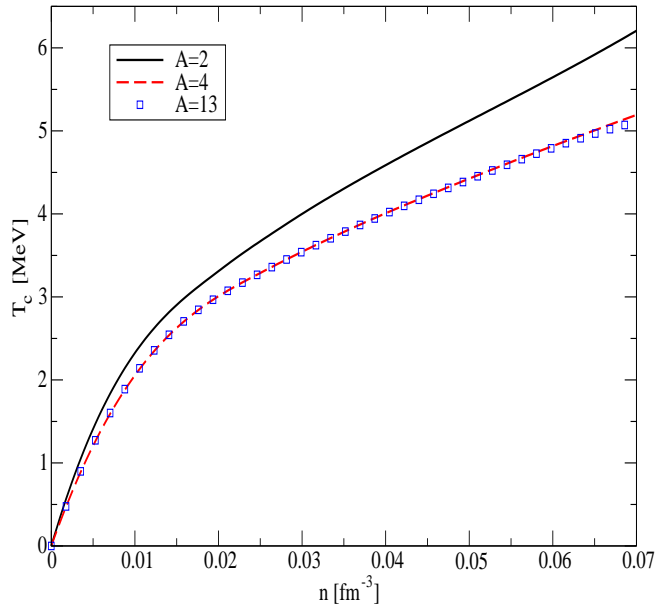}
\vskip 0.3cm
\caption{(Color online)
Dependence of critical temperature of Bose-Einstein condensation 
of deuterons on matter density for $A\le 2$ matter (solid line, black online)
$A\le 4$ matter (dashed line, red online), and $A\le 13$ matter (squares).
}\label{fig4}
\end{figure}
\begin{figure}[!] 
\vskip .6cm
\psfig{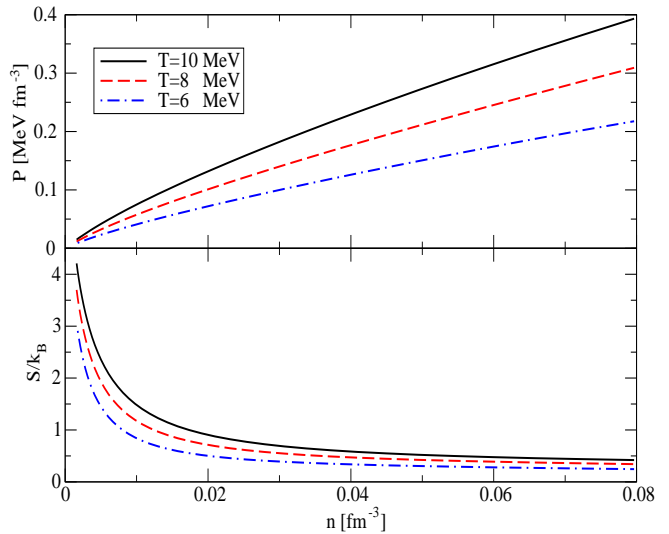}
\vskip .3cm
\caption{(Color online)
Dependence of pressure (upper panel) and entropy (lower panel) 
on the density of matter for temperatures 
$T=10$ MeV (solid line, black online), 
$T=8$ MeV (dashed line, red online), and 
$T=6$ MeV (dashed-dotted line, blue online). $k_B$ is the Boltzmann 
constant.
}
\label{fig5}
\end{figure}
Figure.~\ref{fig3} shows the chemical potentials of species under 
the conditions discussed in Fig.~\ref{fig1}. The relative ordering
of the chemical potentials follows from Eq.~(\ref{eq:cheq}).
Because $\mu_n\simeq\mu_p$, their absolute value scales as 
$\mu_{\alpha}\sim A\mu_n$. The negative sign of chemical potentials 
of bosonic (integer total spin) nuclei implies that these are  
above their critical temperature and density of Bose-Einstein 
condensation (see Fig.~\ref{fig4}).  The condition $\mu_{\alpha}(T) 
=0$ at fixed density is first fulfilled for deuterons (the critical 
temperature of Bose condensation scales  as 
$T_c\sim M^{-1}$, where $M$ is the boson mass). The critical 
temperature of BEC of deuterons as a function 
of density is shown in Fig.~\ref{fig4} in matter consisting of nucleons 
and deuterons only ($A \le 2$), nuclei with mass number $A \le 4$ and 
nuclei with mass number $A \le 13$.  It is seen that the presence of nuclei 
with $A >2$ reduces the critical temperature of BEC, whereby the effect 
of adding the nuclei with mass number $4 < A \le 13$ has little effect, 
because
their fraction is small. Physically, the presence of heavier nuclei reduces 
the fraction of deuterons in matter, i.e., the ``effective'' density of 
deuterons. Therefore, the critical temperature $T_c \sim n_d^{2/3}$ 
for BEC is reduced.

Figure~\ref{fig5} displays the pressure and entropy as a function 
of density for several constant temperatures. It is seen that the 
pressure is large for large temperatures and increases linearly 
with density. The entropy is largest at low densities and decreases 
rapidly as the density is increased. It is seen that it scales 
linearly with temperature, as one would expect for degenerate fermionic 
matter. A comparison of our equation of state with those of 
Refs.~\cite{Lattimer:1991nc} and \cite{Shen:1998gq} shows that the differences 
are small, as expected.

\begin{figure}[t] 
\vskip 1cm
\psfig{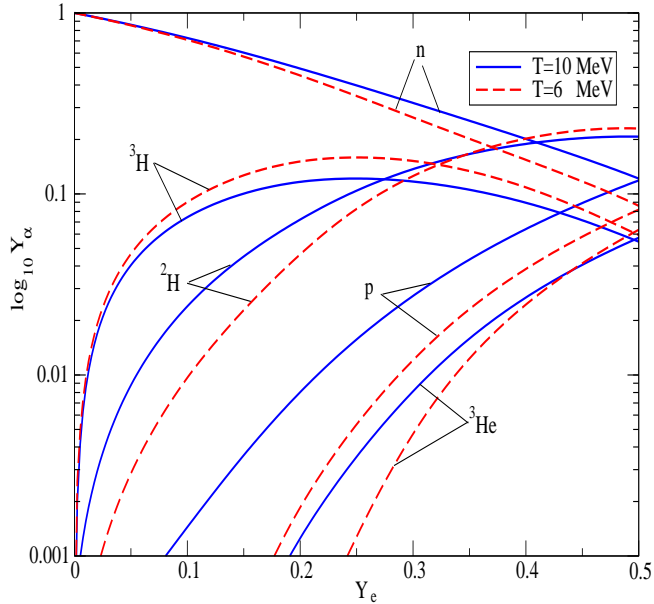}
\vskip 0.4cm
\caption{(Color online) Dependence of abundances of nuclei in matter with 
$A\le 4$ on electron fraction $Y_e$ for fixed density 
$n = 0.041$ fm$^{-3}$ and two temperatures  
$T=10$ MeV (solid lines, blue online) and 
$T=6$ MeV (dashed lines, red online).
}\label{fig6}
\end{figure}
\begin{figure}[t] 
\vskip 1cm
\psfig{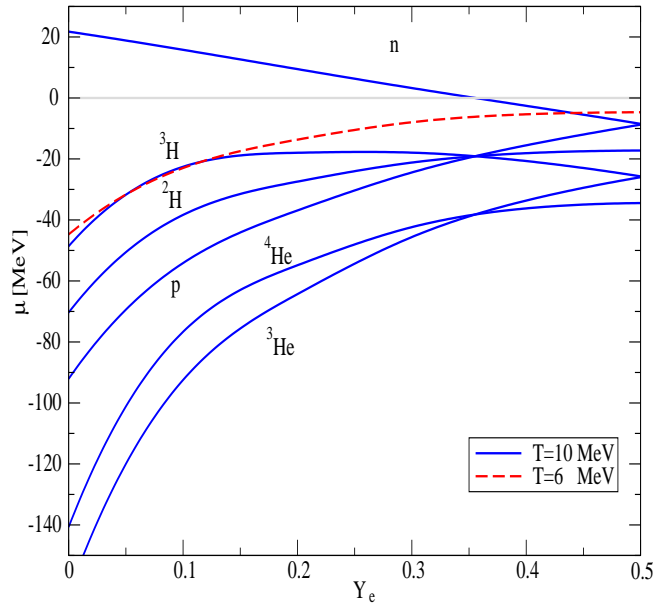}
\vskip 0.4cm
\caption{(Color online)
Dependence of the chemical potentials of nuclei in matter with 
$A\le 4$ on electron fraction $Y_e$ for fixed density $n = 0.041$ 
fm$^{-3}$ and temperature $T=10$ MeV (solid lines, blue online). 
The deuteron chemical potential is shown also at $T=6$ MeV 
(dashed line, red online).
}\label{fig7}
\end{figure}
\begin{figure}[!] 
\vskip 1. cm
\begin{center}
\psfig{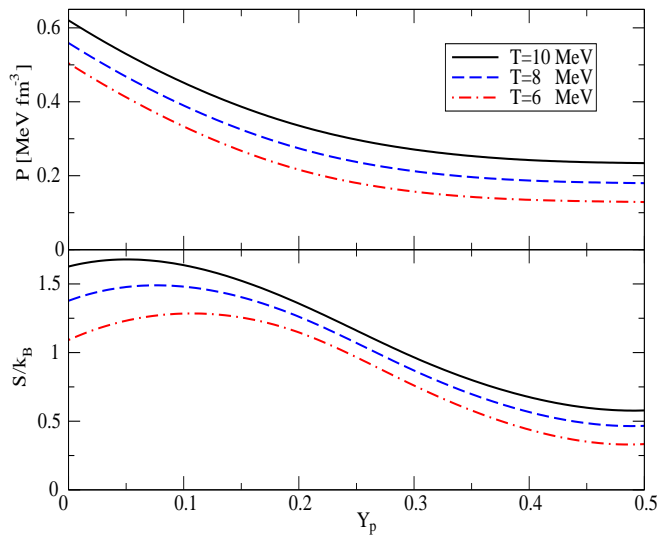}
\end{center}
\caption{(Color online)
Dependence of pressure (upper panel) and entropy (lower panel) 
on electron fraction $Y_e$ at density $n = 4.1 \times 10^{-2}$ 
fm$^{-3}$ and several temperatures.
}\label{fig8}
\end{figure}
\begin{figure}[!] 
\begin{center}
\psfig{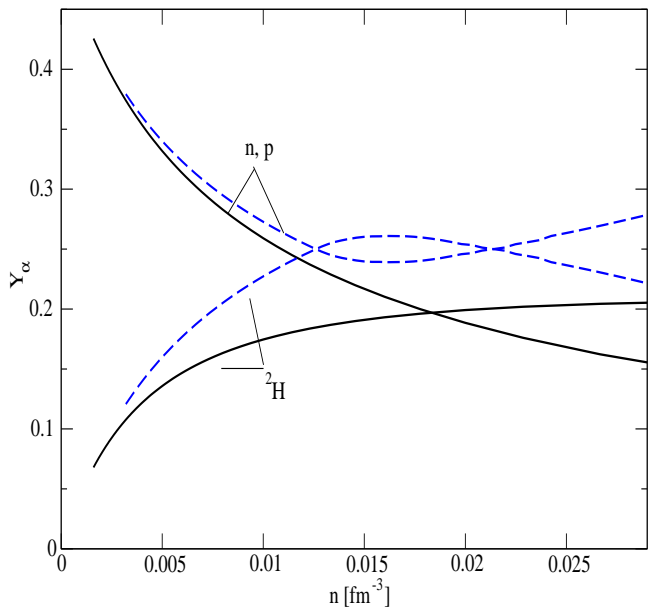}
\end{center}
\caption{(Color online)
Dependence of nucleon ($n$ and $p$) and deuteron ($^2$H) abundances
on density of matter at $T=10$ MeV. Results with free space binding 
energies are shown by solid lines (black online), those with medium modified 
deuteron binding energies by dashed lines (blue online).}
\label{fig9}
\end{figure}
\subsection{Isospin asymmetric matter ($\chi\neq 0$)}
In this subsection we study light nuclei in isospin asymmetric 
nuclear matter. We consider proton-deficient matter, i.e.,
$0\le \chi = (n_n-n_p)/n\le 1$, or in terms of electron fraction
$0\le Y_e\le 0.5$, which is the relevant case in 
supernovas and neutron stars. The dependence of the 
abundances of light nuclei on the electron fraction $Y_e$ at
fixed density $n=0.041$ fm$^{-3}$ and two temperatures $T=10$ MeV 
and $T=6$ MeV is shown in Fig. \ref{fig6}. 
Consider first deuterons 
(the arguments below apply equally to $\alpha$ particles and other 
nuclei with equal numbers of protons and neutrons). Their abundance 
is maximal for $Y_e=0.5$. Increasing asymmetry reduces the number 
of protons that are available for building a deuteron; consequently 
the number of deuterons reduces with increasing asymmetry and in 
the limit $Y_e = 0$ they are extinct. Asymmetry breaks the degeneracy 
between the abundances of $^3$H and $^3$He; the abundance of $^3$He, 
which requires two protons per neutron, decreases most rapidly. 
The abundance of triton ($^3$H) is nonmonotonic: it first increases 
because excess neutrons can be easily accommodated in nuclei and then 
decreases because the number of available protons vanishes. These two 
effects make a compromise when $Y_e\simeq 0.25$, where triton 
abundance is maximal. Note that the ratio of abundances of deuterons to 
tritons is inverted for large asymmetries; indeed, in symmetric nuclear 
matter the deuterons are the second most abundant species, while in asymmetric 
matter their abundances fall below those of tritons for large enough 
asymmetries. Lower temperatures are seen to increase the proton depletion,  
$^3$He and deuteron abundances decrease faster, and the increase 
in triton abundance at $Y_e \le  0.5$ is more pronounced.

The dependence of chemical potentials of light nuclei on 
electron fraction at fixed density $n=0.041$ fm$^{-3}$ 
and temperature $T=10$ MeV is shown in Fig.~\ref{fig7}. 
The behavior of chemical potentials 
is  understood in analogy to the behavior of abundances discussed 
above: the nuclei with equal neutron to proton ratio, as well 
as those that require proton excess are disfavored by asymmetry and
their chemical potentials are negative and large. The chemical 
potentials of tritons ($^3$H) are nonmonotonic functions of 
$Y_e$ because excess neutrons are responsible for its increase 
for $Y_e \le 0.5$, while the proton extinction for $Y_e \ge 0$ 
is responsible for its decrease.

The dependence of pressure and the entropy on the electron fraction $Y_e$ 
is shown in Fig.~\ref{fig8}. It is seen that the pressure is lowest in 
the symmetric case and increases with the asymmetry; like in the 
symmetric case larger temperatures sustain larger pressures and entropies. 
The entropy increases with asymmetry starting from the 
neutron matter limit $Y_e=0$, an increase associated with the onset 
of new degrees of freedom (nuclei), which is followed by a decrease
as one approaches the isospin symmetric limit.

\section{Summary and outlook}
\label{sec:conclusions}

In this article we set up a quasiparticle gas formalism to compute the 
equation of state and composition of dilute isospin symmetric and 
asymmetric nuclear matter for applications to supernova physics. Our key 
finding is that matter is dominated by the light nuclei, such as deuterons, 
tritons ($^3$H), and $^3$He isotopes of helium. The $\alpha$-particles
($^4$He) contribute less than $1\%$ to the number density. Furthermore, 
we find that in a large portion of the density and temperature diagram 
deuterons form a Bose-Einstein condensate. The effect of heavier clusters 
is to reduce the critical temperature of Bose-Einstein condensation of 
deuterons. A novel feature of isospin 
asymmetric matter is the enhancement of the abundances of neutron-rich 
nuclei and the corresponding suppression of proton-rich ones. This is clearly 
manifest in the enhancement of triton abundances with increasing asymmetry,
which makes tritons the most abundant species after neutrons in asymmetric
nuclear matter. Compared to isospin symmetric matter the relative abundance
of deuterons and tritons is inverted in strongly asymmetric matter.

The present setup is a useful platform for further extensions of the 
theory, which we would like to discuss briefly. The binding energies of 
light nuclei are generally functions of density and temperature. At high
densities and low temperatures the binding energies are reduced and at 
some critical values of these parameters bound states are dissolved
(see Ref.~\cite{Ropke:2008qk} and references therein). 
Thus, for example, nuclei will disappear in matter at high densities 
leaving behind a uniform nuclear fluid. The critical extinction line for 
deuterons and tritons in the phase diagram of symmetric nuclear matter
was obtained recently in Ref.~\cite{Sedrakian:2005db}. 
In  Fig.~\ref{fig9} we show the effect of incorporating the 
temperature-density-dependent binding energies
of deuterons, computed in Ref.~\cite{Sedrakian:2005db}, on the composition 
of matter with mass numbers $A\le 2$. It is seen that the high-density 
asymptotic state of abundances is inverted; the abundances of deuterons
are larger than the nucleonic abundances for constant, free space,
binding energies. However, their relative ratios are inverted when 
the reduction of the deuteron binding energies at large densities is 
taken into account. Matter effects will affect the abundances of other
light nuclei in a similar way, which will guarantee that the high-density 
asymptotic state corresponding to the continuum of nuclear fluid at 
saturation density is recovered. 

Apart from the statistical effect of suppression of bound state energies in 
matter further aspects that should be incorporated in the model include
(i) leptons and electromagnetic forces  (screening 
of nuclear charge), (ii) onset of $\beta$ equilibrium during the late 
time dynamics of supernovas, (iii) elastic scattering among the light 
nuclei themselves and with nucleons, and (iv) reactions. Of course, 
larger numbers of nuclei and resonances (nuclei with short-decay 
times scales) can be easily incorporated within our model.

This work was in part supported by the Deutsche Forschungsgemeinschaft
(Grant SE 1836/1-1).

\end{document}